\documentclass[print,twocolumn, APS, prl]{revtex4-1}
\usepackage{graphicx}
\usepackage{dcolumn}
\usepackage{bm}
\usepackage{wrapfig}
\usepackage{amssymb}
\usepackage{soul}
\usepackage{amsmath}
\usepackage[breaklinks=true,colorlinks,linkcolor=blue,citecolor=green]{hyperref}
\def\be{\begin{equation}}
\def\ee{\end{equation}}
\def\e#1{\label{#1}\end{equation}}
\def\bea{\begin{eqnarray}}
\def\eea{\end{eqnarray}}
\def\ea#1{\label{#1}\end{eqnarray}}

\def\bem#1{\begin{mathletters}\label{#1}}
\def\eml{\end{mathletters}}

\def\ket#1{{|#1\rangle}}

\def\mean#1{{\langle#1\rangle}}
\def\4#1{{\boldsymbol{#1}}}
\def\8#1{{\widetilde{#1}}}
\def\bse{\begin{subequations}}
\def\ese{\end{subequations}}

\def\Rb87{$^{87}\text{Rb}$}
\def\0{\ket{0}}
\def\1{\ket{1}}

\def\be{\begin{equation}}
\def\ee{\end{equation}}
\def\e#1{\label{#1}\end{equation}}
\def\bea{\begin{eqnarray}}
\def\eea{\end{eqnarray}}
\def\ea#1{\label{#1}\end{eqnarray}}

\def\bem#1{\begin{mathletters}\label{#1}}
\def\eml{\end{mathletters}}

\def\ket#1{{|#1\rangle}}

\def\mean#1{{\langle#1\rangle}}
\def\4#1{{\boldsymbol{#1}}}
\def\8#1{{\widetilde{#1}}}
\def\bse{\begin{subequations}}
\def\ese{\end{subequations}}

\DeclareMathOperator{\sinc}{sinc}
\usepackage{graphicx}
\usepackage{dcolumn}
\usepackage{bm}
\usepackage{wrapfig}
\usepackage{amssymb}
\usepackage{soul}
\usepackage{amsmath}
\begin{document}
\title{Ancilla assisted Discrete Time Crystals in Non-interacting Spin Systems}
\author{Jianpei Geng, Vadim Vorobyov, Durga Dasari and Joerg Wrachtrup }
\address{3.\,Physics Institute, University of Stuttgart, Center for Applied Quantum Technologies, IQST, MPI for Solid State Research, Stuttgart 70569, Germany}
\begin{abstract}
We show here through experiments and exact analytical models the emergence of discrete time translation symmetry {breaking} in non-interacting systems. These time-periodic structures become stable against perturbations only in the presence of their interaction with their ancillary quantum system and display subharmonic response over a range of rotation angle errors. We demonstrate this effect for central spin and spin-mechanical systems, where the ancillary induced interaction among the spins stabilizes the spin dynamics against finite errors. Further, we extend these studies and show the possibility to even achieve non-local (remote) synchronization of such Floquet crystals.

\end{abstract}

\maketitle

\section{Introduction}

Time-periodic self-organized structures in the Hilbert space of a collective many body system \cite{QTC_Wilczek_PRL, AbsQTC_PRL, FTC_PRL, PhaseStrucDQS_PRL, DTCPhaseDiag_PRL, BoundTC_PRL, PrePhaseTTS_PRX, LongRangePreTC_PRX, MBLThermEntang_RMP, DTC_ARCMP, HistoryTC_arxiv1910_10745} have been extensively studied and demonstrated experimentally over the last years for many physical systems \cite{ObservDTC_TrapIon_Nature, ObservDTC_NV_Lukin_Nature, ExpDTC_StarShape_NMR_PRL, ObservDTC_NMR_PRL, ObservQuasiTC_PRL, ObservSTC_SuperfluidQGas_PRL, DTC_CQED_PRL, ObservDissipativeTC_PRL, DTC_DissipativeESR_NJP, FPre_NMR_NaturePhys, ProgramDTC_NISQ_arxiv2007_11602, DTC_SupercondQC_arxiv2105_06632, DTC_ProgNV_arxiv2107_00736, FPrePhase_SC_Pan_arxiv2107_07311, ObservPreDTC_TrapIon_Science}. Time crystals {(TCs)} \cite{QTC_Wilczek_PRL, FTC_PRL, DTC_ARCMP, HistoryTC_arxiv1910_10745} have been mostly studied as a non-equilibrium phase emerging due to the spontaneous breaking of time translational symmetry (TTS). 
Driven/Floquet quantum systems exhibit a discrete time-translation symmetry (dTTS) {in the Hamiltonian}. 
It was shown that this symmetry can be broken in a Floquet time-crystal \cite{FTC_PRL, PhaseStrucDQS_PRL}, leading to a “subharmonic” response of quantum observables that oscillate with an
integer multiple of the driving period.
{The subharmonic response should be robust against weak perturbations in the driving field to exhibit a generic discrete time crystal (DTC) phase.}
The interplay of {periodic driving, disorder, and interaction plays an important role in stablizing the DTC phase, which}
has been studied extensively for Ising and dipolar spin systems \cite{FTC_PRL, PhaseStrucDQS_PRL, DTCPhaseDiag_PRL, PrePhaseTTS_PRX, LongRangePreTC_PRX, DTC_ARCMP, HistoryTC_arxiv1910_10745}.
{Experimentally these effects have been observed in trapped ion system \cite{ObservDTC_TrapIon_Nature, ObservPreDTC_TrapIon_Science}, ensemble nitrogen-vacancy center electron spins in diamond \cite{ObservDTC_NV_Lukin_Nature}, NMR platforms \cite{ExpDTC_StarShape_NMR_PRL, ObservDTC_NMR_PRL}, and ultracold atoms \cite{ObservQuasiTC_PRL, ObservSTC_SuperfluidQGas_PRL, ObservDissipativeTC_PRL}, and have been simulated with programmable quantum simulators \cite{DTC_SupercondQC_arxiv2105_06632, DTC_ProgNV_arxiv2107_00736}}
Here we extend these studies to even non-interacting systems, and show how an ancilla mediated coupling could serve to establish 
a {DTC} behavior.
We would like to analytically show and experimentally demonstrate how ancilla assisted 
non-interacting quantum systems could still be a test bed to observe {the DTC signature}. 
Further, we discuss how {it} 
can be extended to achieve  remote/non-local stabilization of the DTC-phases.

In this work, we will consider the non-interacting spins interacting with a single ancillary system that can either be a well-controlled electron spin or a cantilever/cavity mode. By periodically alternating the spin rotations and the ancilla interactions, we find that a {subharmonic} periodic phase emerges that is robust against rotation angle errors. Deviating from previous works, we first show {the stabilization of the DTC signature by} the ancilla induced interactions among the non-interacting spins, and then show an experimental demonstration of our model {with electron and nuclear spins of nitrogen-vacancy (NV) center in diamond}.
We will consider three scenarios: (i) {DTC in non-interacting nuclear spins mediated by an ancillary central electron spin,} 
(ii) Phonon mediated DTC among distant electron spins and (iii) Stabilization of two remote non-interacting spin ensembles through their coupled ancillary spins. 

\section{Electron-Nuclear spin system}
We will consider a central spin system  where the central electron spin is coupled to N non-interacting nuclear spins. The Hamiltonian describing their dynamics is given by
\be
H=S^z\sum_k g_k I^z_k + \omega S^x,
\ee
where $g_k$'s are the respective couplings to the nuclear spins and $\omega$ is the driving field for the central spin. Here we would like to highlight the role of the additional driving field that makes the Hamiltonian non-trivial. In the absence of such a field, there is no induced interaction among the nuclear spins. On the contrary for $\omega \ne 0$, one can generate entanglement among the nuclear spins. The Hamiltonian here can again be solved exactly and the Floquet dynamics generated by the operator $U_F$ can be solved exactly for imperfect driving of the nuclear spins by an angle $\theta = \pi - \epsilon$. As the drive is identical for all the nuclear spins, and further assuming $g_k = g$, we can solve the dynamics in the total angular momentum basis of the nuclear spins when initialized in the state $\ket{N/2}=\ket{\uparrow\uparrow\cdots\uparrow}$. This state belong to the collective subspace of $I = N/2$, and the dynamics does not allow for mixing of the collective spin sub-spaces when $g_k = g$. 
\begin{figure*}
\begin{center}
\includegraphics[width=1.0\textwidth]{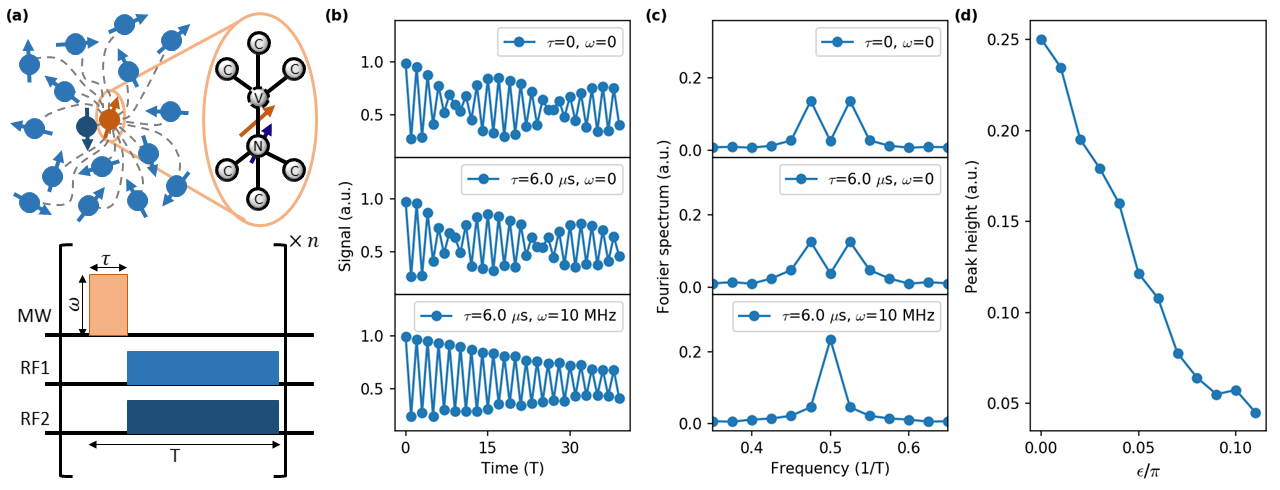}
\end{center}
\label{level}
\vspace{-5mm}
\caption{{\bf DTC in Central spin systems}. {(a) {Scheme for observation of ancilla induced DTC signature based on NV center in diamond. The upper panel shows schematic representation of the system. The NV center electron spin (orange) plays the role of the ancilla and couples to several $^{13}$C nuclear spins (blue). The $^{13}$C nuclear spin with strongest coupling of $414$ kHz to the electron spin is represented with dark blue arrow. The lower panel shows the pulse sequence for the DTC experiment. The MW pulse of duration $\tau$ resonantly drives the electron spin with $\omega S_x$ term in the Hamiltonian. The nuclear spins are driven by two RF-pulses to correct for the ancillary spin-state dependent hyperfine coupling. 
(b) The time-domain response of the nuclear spin. 
Only with the driving MW field ($\tau\neq0, \omega\neq0$) it shows subharmonic response. The rotation error is set as $\epsilon=0.05\pi$.
(c) FFT spectra of b. {The FFT peak splits into two due to the rotation error $\epsilon$ without the driving MW field ($\tau = 0$ or $\omega = 0$), while it remains at 0.5 with the driving MW field ($\tau\neq0, \omega\neq0$)} (d) Spectral peak height of DTC phase as a function of $\pi$ pulse error.}}}
\end{figure*}
We now implement the DTC protocol for the non-interacting spins, i.e., we alternate the evolution of the spins by the two unitary opeartor viz., the rotation of spins along the x-direction i.e., $U_R(\theta)  = \exp(i\theta\sum_k I^x_k)$ and the evolution $U(t)\equiv  \exp(itH)$ (see Supp. Info for exact solution), thus realizing the stroboscopic Floquet evolution governed by the combined operator
\be
U_F = U_R(\theta) \cdot U(g,\omega,\tau)
\ee
We start with an initial uncorrelated state of the spins, which is an eigenstate of the operator $L_z$. Clearly, the spin-mechanical evolution has no influence on such a state. Further, if the rotation angle $\theta$ is chosen such that $\theta = \pi$, then no superposition states are generated and the initial state evolves within the two-dimensional subspace of the $L_z$ eigenstates. Thus the combined evolution $U_F$ can be seen as a flip between two states. Further if the evolution due to both the terms are equally spaced in time we immediately see a subharmonic Fourier response at half the drive frequency, as the spins remain idle during their evolution with the electron spin.

In the absence of the spin-spin coupling any imperfection in the rotation angle $\theta = \pi-\epsilon$ immediately destroys the subharmonic response. To see this, let us consider the example of $N$-spins initialized in the state $\ket{m} = \ket{\uparrow\uparrow\cdots\uparrow}$, where $m = \sum_k\mean{I^z_k}$, represents the total magnetization of the spins. When the rotation angle $\theta = \pi$, one immediately finds that the spins are rotated from $\ket{m} \rightarrow \ket{-m}$. In the presence of an error $\epsilon \ll 1$, one can find that the rotation leads to mixing of different $m$ subspaces i.e.,
\be
\ket{m} \rightarrow \ket{-m} + \epsilon\ket{1-m} + \mathcal{O}(\epsilon^2) + \cdots
\ee

Now keeping only terms linear in, $\epsilon$, and by rotating the spins repeatedly by the misaligned angle $\theta$, one finds after $M$(even) number of rotations, that the spins accumulate an error that scales as
\be
\ket{m} \rightarrow \ket{m} + \frac{M\epsilon}{2}(\ket{1-m} + \ket{m-1})+ \mathcal{O}(\epsilon^2).
\ee
Due to this mixing of the unwanted subspaces, one immediately finds that the sub-harmonic response is destroyed and the central peak is split into satellite peaks whose strength and spacing is determined by the error angle $\epsilon$. 

We now turn-on the spin-spin interaction $U(g)$ for a time $t$, the mixing now differs dramatically from the non-interacting case and is given by,
\be
\ket{N/2} \rightarrow \ket{N/2} + \epsilon[f_1(M)\ket{N/2-1} + f_2(M)\ket{1-N/2}].
\ee
Here $f_1(M)=Tr(A_M\rho_S(0))$ and $f_2(M)=Tr(B_M\rho_S(0))$, and the operators $A_M, ~B_M$ are multiples of central spin unitaries (see Suppl.Info). The factors $f_{1,2}$ shown non-linear response as a function of $\epsilon$ and the Floquet cycle number $M$ that leads to a suprression of the satellite peaks, and the resurrection of the central subharmonic peak as can be seen in Fig. 2. Here we would also like to highlight again the role of $\omega$, in whose absence the factors $f_1, f_2$, become linearly dependent on the flip-number $M$ as shown in Eq. (5). For identical couplings to the central spin one could numerically simulate the effects of Floquet stabilization in large spin-baths exceeding $N > 10^3$ (see Supp. Info).  We now experimentally verify this effect for the central spin system formed by the solid-state spins in diamond, where the electron spin of the NV center forms the central spin mediating the interaction between the non-interacting ${}^{13}C$-nuclear spins.

\section{Experiments}
The experiments were carried out at room temperature on a home-built confocal microscope setup using a type IIa CVD grown diamond crystal (layer) that has [100] surface orientation and a $^{13}$C concentration of $0.2\%$ \cite{Waldher2014}. The NV center is located roughly $15 \, \mathrm{\mu m}$ below the diamond surface and a solid immersion lens has been carved around it via focused ion beam milling. An omega-loop waveguide for microwave (MW) and RF excitation is fabricated onto the diamond via optical lithography. An external static magnetic field of $\sim 650$mT from a permanent magnet is aligned along the symmetry axis of the NV centre (z-axis). A 520 nm laser light is focused onto the NV by an oil immersion objective, which also collects light from the diamond. The fluorescence light of the NV center is isolated by spatial and spectral filtering and finally detected with a single photon counting detector (APD Perkin Elmer). Optical excitation of the NV centre polarizes the electron spin triplet (S=1) into its $m_S = 0$ state, which forms the source of initial spin polarization in the experiment.

In our experiments we use $^{13}C$ nuclear spin-bath {in which six nuclear spins are well resolved} \cite{Zaiser2016, QFT_Vadim}, and the NV center electron spin plays the role of the ancillary spin (see Fig. 1). As the ancillary spin is a three-level system we choose the two-level subspace spanned by the state $\ket{m_S=0}$, and $\ket{m_S=-1}$. The ancillary spin couples strongly to its intrinsic $^{14}N$ spin and to a closest $^{13}C$ with strength$A_{zz} = 2 \pi  \cdot 414$  kHz. We use this spin for observing the field-stabilized Floquet behavior in the presence of errors. There is an always-on interaction between the electron and nuclear spins. 
We start by initializing the ancilla, the $^{14}$N nuclear spin, and the most strongly coupled $^{13}$C nuclear spin into a fully polarized state. The MW pulse of duration $\tau$ resonantly drives the electron spin with $\omega S_x$ term in the Hamiltonian. The rotations of the nuclear spins are  realized by two RF pulses with two frequencies $7.126 MHz$, and $7.539 MHz$ simultaneously. The first frequency equals the Larmor frequency of $^{13}$C nuclear spins, drives the weakly coupled $^{13}$C nuclear spins and the most strongly coupled $^{13}$C nuclear spin when the ancillary electron spin in the state $\ket{m_S=0}$, and the second RF pulse with frequency of 7.539 MHz drives the most strongly coupled $^{13}$C nuclear spin when the ancilla is in the state $\ket{m_S=-1}$. As the ancilla spins is continuously rotating due to its microwave field $\omega$, toacheive resonant nuclear spins independent of the ancilla spin state we apply the two $RF$-fields as described above. Now we make these rotations imperfect by driving them at angles of $\pi-\epsilon$ periodically. After implementation of the sequence shown in Fig. 1, the DTC phase gets stabilized for all the nuclear spins, and to confirm we only measure here the strongly coupled $^{13}$C nuclear spin using single-shot readout. 

We compare the cases when the driving field $\omega$ is switched on and off over a time-scale $\tau$ that determines the free evolution $U_g(t)$ (see Eq.(2)). 
In between the stroboscopic evolution we perform incomplete flipping of the nuclear spins.
This evolution is sandwiched between the local rotation $\theta$ operations of the nuclear spins to observe the emergence of subharmonic response even for imperfect driving as shown in Fig. 1b. 
{By adding the interaction time period $\tau$ enabled by driving field $\omega$ we observe the stabilization of subharmonic response in the presence of the errors (Se fig 1b,c). The stabilization we observe is stable, but the relative peak height is decaying with increase of the nuclear spin driving errors as was also shown for directly interacting systems \cite{ObservDTC_TrapIon_Nature, ObservDTC_NV_Lukin_Nature, ExpDTC_StarShape_NMR_PRL, ObservDTC_NMR_PRL}.}

Below we extend our analysis to its full generality by considering another class of ancilla-assisted systems such as spin-mechanical systems where the induced interaction becomes clearly visible in the evolution operator in the form of Ising Hamiltonian \cite{DTCPhaseDiag_PRL}, thereby retaining temporal order even in the absence of any external driving.


\section{Spin-Mechanical system}
Here we will consider a spin-mechanical system realized through single spins implanted in a nanomechanical oscillator (NMO). In the presence of strain or  a large magnetic field gradient, these spins (two-level systems) couple to a position-dependent magnetic field i.e., the spins experience a phase shift of their energy states that depends on the position of the cantilever. The coupling between single spin and a given mechanical mode of frequency $\omega_m$ is determined by the zero point motion of the oscillator and the magnetic field gradient \cite{tommy}. The Hamiltonian describing the dispersive interaction between the spin and the NMO (in units of $\hbar = 1$) is given by
\be
\hspace{-5mm}
H=\omega_0 a^\dagger a + \sum_k g_kS^z_k (a +  a^\dagger)] + \Omega(t)S^x.
\ee
where $g_k$ is the coupling between the k-th spin ($S$) and the (ancillary) mechanical mode of the oscillator ($a_m$) that has a frequency $\omega$ The spin can further be time-dependent fields $\Omega(t)$. In addition to this unitary coupling the spins and the oscillator modes suffer nonunitary decay processes which we will consider in the later part of this paper.
In the absence of any external control ($\Omega(t) = 0$), the time evolution due to the above can be solved exactly \cite{durgaprl}, and it closed-form solution is given by
\be
U(t) = {\rm e}^{-itf(t)L_z^2}\exp\left[L_z\left(\alpha_0a^\dagger-\alpha^*_0a\right)\right]\ee

where $L_z = \sum_k g_{k} S^z_k$ and the time-dependent couplings among the spins induced by the NMO are $f(t)=\frac{1}{\omega_0}(1- {\sinc} (\omega_0t))$, and the spin decoherence due to the cantilever is given by the second exponential term with $\alpha_0 =  \frac{1-{\rm e}^{i\omega_0t}}{\omega_0}$. One can clearly see from Eq. (2) the induced interaction among the non-interacting spins.

\begin{figure}
\includegraphics[width=0.5\textwidth]{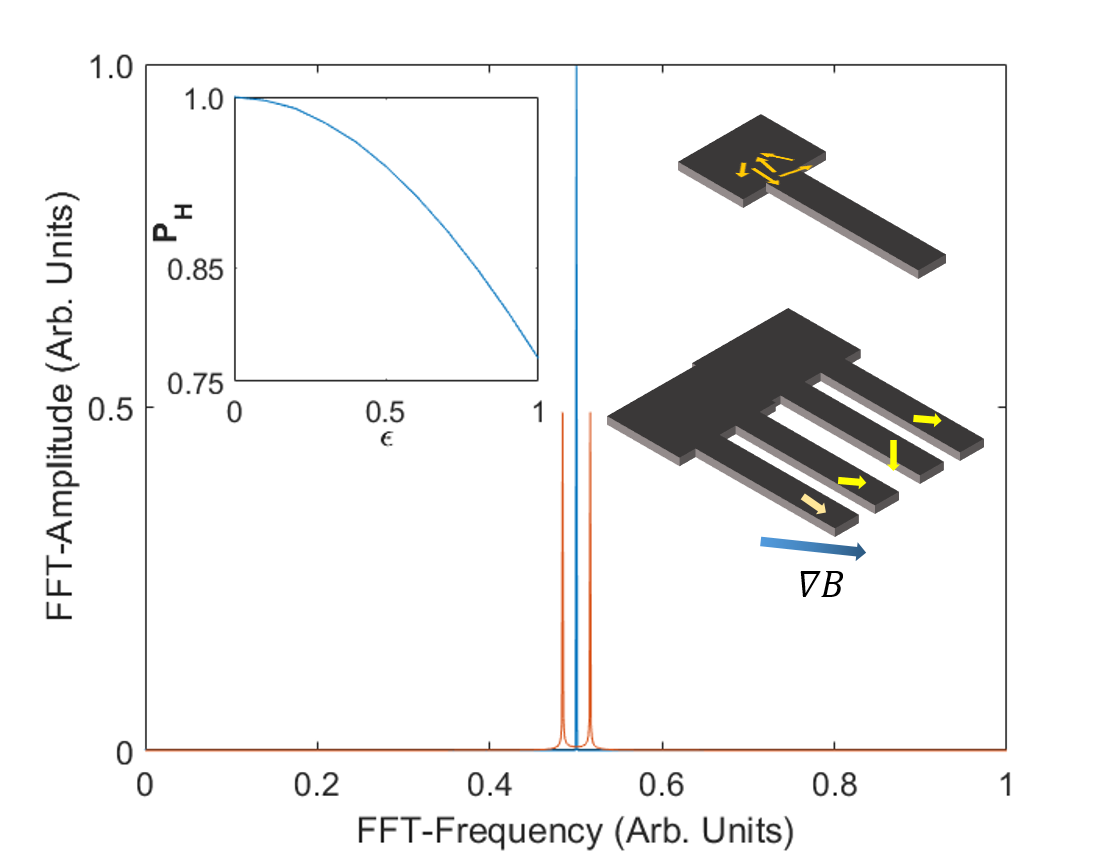}
\caption{{\bf DTC in Spin-mechanical system}. We show here the subharmonic response of the electron spins in the presence and absence of the coupling to its common mechanical mode. Inset1: Schematic representation of the spin-mechanical system realized by micro/nano-mechanical cantilevers with spins embedded in it. Inset2: The subharmonic peak height as a function of the rotation error $\epsilon$ in the case of spin-mechanical coupling.}
\end{figure}

In the absence of the spin-spin coupling any imperfection in the rotation angle $\theta = \pi-\epsilon$ immediately destroys the subharmonic response, as shown earlier in Eq. (3).
We now turn-on the spin-mechanical interaction $U(g)$ for a time $t$. To obtain a closed form solution we consider identical couplings between the spins and the NMO i.e., $g_k = g$ \cite{durgaprl}. In this case the spin-interaction term in Eq. (2) simply represents the Squeezing operator. We now stroboscopically switch on this interaction between any two consecutive rotations of the spins. In the presence of this non-linear interaction the rotation now stabilizes as the leakage to the unwanted subspaces is dramatically reduced as (see Supp. Info)
\be
\ket{m} \rightarrow \ket{m} + \frac{\epsilon \sin^2[Ngt(1+M)]}{\sin^2[Ngt]}(\ket{1-m} + \ket{m-1})+ \mathcal{O}(\epsilon^2).
\ee
From the above equation one could see that the mixing to unwanted subspaces does not scale linearly with the error angle but is rather an oscillatory function whose amplitude for $Ngt=\pi$ is bounded by unity, for any value of $M$. Due to this we regain our subharmonic response (see Fig. 2) indicating the robustness of the spin rotations to weak perturbations in the presence of an ancillary system. While the above analytical treatment is valid for very weak perturbations, in general the the rigidity can be numerically simulated for a range of $\epsilon$ where the location of the normalized Fourier peak remains locked precisely at $\omega = 1/2T $.
\begin{figure}
\includegraphics[width=0.5\textwidth]{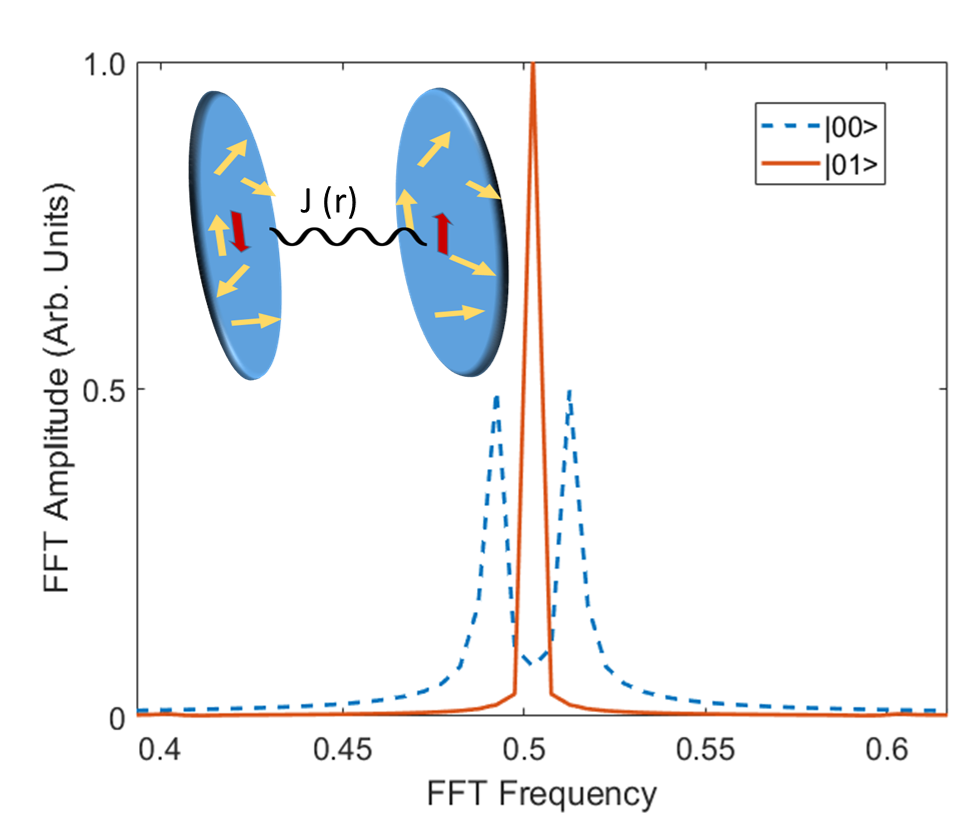}
\caption{{\bf Remote Synchronization of DTC's.} Two electron spins (red arrows) couple to each other with strength $J$, and to their local nuclear spin bath (yellow arrows). We show how the coupling between the electron spins synchronizes the DTC signature of the local nuclear spin baths dependent on the state of the ancillary spins. When the ancillary spins are in state $\ket{00}$, the $J-$ coupling has no role, and hence the nuclear spins show non-DTC signature. On the otherhand when they are in state $\ket{01}$ the $J$-coupling flip-flops the spin states thereby correcting the errors and resurrecting the DTC signature at both the remote sites simultaneously.}
\end{figure}
\section{Remote Synchronization of DTC's}
We now consider a novel scenario to show how DTC in different spatial locations can be synchronized by coupling the central electron spins from each location (see Fig. 3). This is a relevant scenario for solid-state spin systems where the electron spin associated to the defect centers are coupled to their local nuclear spin-bath and can also be dipolar coupled to a nearby defect electron spin that is coupled to its local nuclear-spin bath \cite{nvnv, EntangSwapOptimalCont_NV_NC}. The Hamiltonian describing such a system is given by
\be
H=S^z_1\sum_k g_{k,1} I^z_{k,1} + S^z_2\sum_k g_{k,2} I^z_{k,2} + J({S}^+_1{S}^-_2+{S}^-_1{S}^+_2).
\ee
Unlike the earlier scenario, we do not use any driving field for the central spins. In fact, the coupling between the $S$-spins marks the  onset of the spin oscillations if they are initialized in the correct state. To see this we again start with two possible initial states of the central spins. When initialized in the state $\ket{00}$, the $J$-coupling has no influence as it is an eigenstate of the coupling term. On the other hand when initialized in the state $\ket{01}$, the coupling term mixes continuously with the state $\ket{10}$ at a rate $J$. Such a mixing at a rate $J$ is good enough to synchronize both the remote nuclear spin-bathes such that they show collectively the sub-harmonic response and its stability to weak perturbations of the driving field.

In conclusion we have shown here how a non-interacting spin system can display the features of Discrete Time Crystal in the presence of an ancillary spin that mediates both the interaction among the spins and also modulates their couplings in a fashion that allows the non-interacting spins to regain the subharmonic response even in the presence of induced rotation angle errors. Using single nuclear spins and a single well-controlled ancillary spin we have shown that only in the presence of both the (spin-ancilla) interaction and the driving of the electron spin one can achieve the subharmonic response that is stable even for large rotation errors. We have further generalized these effects to spin-mechanical and remotely connected systems opening new frontiers for Quantum control of solid-state spins in the presence of time-dependent control. Making the electron-nuclear spin interaction dissipative, our central spin model could naturally realize dissipative time crystals observed recently in atom-cavity systems \cite{ObservDissipativeTC_PRL}. We envisage that our results can be further extended in the context of Quantum Thermodynamics where fluctuations in local-temperature can be stabilized by non-local heat exchange between two thermal machines, allowing us to even crystallize entropy both in space and time.

\begin{acknowledgements}
We would like to acknowledge the financial support by the DFG (FOR2724), ERC project SMeL, 
DFG SFB/TR21, EU ASTERIQS, QIA, Max Planck Society, the Volkswagenstiftung, the Baden-
Wuerttemberg Foundation
\end{acknowledgements}


\begin{thebibliography}{0}%
\makeatletter
\providecommand \@ifxundefined [1]{%
 \@ifx{#1\undefined}
}%
\providecommand \@ifnum [1]{%
 \ifnum #1\expandafter \@firstoftwo
 \else \expandafter \@secondoftwo
 \fi
}%
\providecommand \@ifx [1]{%
 \ifx #1\expandafter \@firstoftwo
 \else \expandafter \@secondoftwo
 \fi
}%
\providecommand \natexlab [1]{#1}%
\providecommand \enquote  [1]{``#1''}%
\providecommand \bibnamefont  [1]{#1}%
\providecommand \bibfnamefont [1]{#1}%
\providecommand \citenamefont [1]{#1}%
\providecommand \href@noop [0]{\@secondoftwo}%
\providecommand \href [0]{\begingroup \@sanitize@url \@href}%
\providecommand \@href[1]{\@@startlink{#1}\@@href}%
\providecommand \@@href[1]{\endgroup#1\@@endlink}%
\providecommand \@sanitize@url [0]{\catcode `\\12\catcode `\$12\catcode
  `\&12\catcode `\#12\catcode `\^12\catcode `\_12\catcode `\%12\relax}%
\providecommand \@@startlink[1]{}%
\providecommand \@@endlink[0]{}%
\providecommand \url  [0]{\begingroup\@sanitize@url \@url }%
\providecommand \@url [1]{\endgroup\@href {#1}{\urlprefix }}%
\providecommand \urlprefix  [0]{URL }%
\providecommand \Eprint [0]{\href }%
\providecommand \doibase [0]{http://dx.doi.org/}%
\providecommand \selectlanguage [0]{\@gobble}%
\providecommand \bibinfo  [0]{\@secondoftwo}%
\providecommand \bibfield  [0]{\@secondoftwo}%
\providecommand \translation [1]{[#1]}%
\providecommand \BibitemOpen [0]{}%
\providecommand \bibitemStop [0]{}%
\providecommand \bibitemNoStop [0]{.\EOS\space}%
\providecommand \EOS [0]{\spacefactor3000\relax}%
\providecommand \BibitemShut  [1]{\csname bibitem#1\endcsname}%
\let\auto@bib@innerbib\@empty
\end{thebibliography}%


\begin{thebibliography}{}
\bibitem{QTC_Wilczek_PRL} F. Wilczek, Quantum Time Crystals, Phys. Rev. Lett. 109, 160401 (2012).
\bibitem{AbsQTC_PRL} H. Watanabe and M. Oshikawa, Absence of Quantum Time Crystals, Phys. Rev. Lett. 114, 251603 (2015).
\bibitem{FTC_PRL} D. V. Else, B. Bauer, and C. Nayak, Floquet Time Crystals, Phys. Rev. Lett. 117, 090402 (2016).
\bibitem{PhaseStrucDQS_PRL} V. Khemani, A. Lazarides, R. Moessner, and S. L. Sondhi,
Phase Structure of Driven Quantum Systems, Phys. Rev. Lett. 116, 250401 (2016).
\bibitem{DTCPhaseDiag_PRL} N. Y. Yao, A. C. Potter, I.-D. Potirniche, and A. Vishwanath, Discrete Time Crystals: Rigidity, Criticality, and Realizations, Phys. Rev. Lett. 118, 030401 (2017).
\bibitem{BoundTC_PRL} F. Iemini, A. Russomanno, J. Keeling, M. Schirò, M. Dalmonte, and R. Fazio, Boundary Time Crystals, Phys. Rev. Lett. 121, 035301 (2018).
\bibitem{PrePhaseTTS_PRX} D. V. Else, B. Bauer, and C. Nayak, Prethermal Phases of Matter Protected by Time-Translation Symmetry, Phys. Rev. X 7, 011026 (2017).
\bibitem{LongRangePreTC_PRX} F. Machado, D. V. Else, G. D. Kahanamoku-Meyer, C. Nayak, and N. Y. Yao, Long-Range Prethermal Phases of Nonequilibrium Matter, Phys. Rev. X 10, 011043 (2020).
\bibitem{MBLThermEntang_RMP} D. A. Abanin, E. Altman, I. Bloch, and M. Serbyn, Many-Body Localization, Thermalization, and Entanglement, Rev. Mod. Phys. 91, 021001 (2019).
\bibitem{DTC_ARCMP} D. V. Else, C. Monroe, C. Nayak, and N. Y. Yao, Discrete Time Crystals, Annu. Rev. Condens. Matter Phys. 11, 467 (2020).
\bibitem{HistoryTC_arxiv1910_10745} V. Khemani, R. Moessner, and S. L. Sondhi, A Brief History of Time Crystals, arXiv preprint arXiv:1910.10745 (2019).
\bibitem{ObservDTC_TrapIon_Nature} J. Zhang, P. W. Hess, A. Kyprianidis, P. Becker, A. Lee, J. Smith, G. Pagano, I. D. Potirniche, A. C. Potter, A. Vishwanath, N. Y. Yao, and C. Monroe, Observation of a Discrete Time Crystal, Nature (London) 543, 217 (2017).
\bibitem{ObservDTC_NV_Lukin_Nature} S. Choi, J. Choi, R. Landig, G. Kucsko, H. Zhou, J. Isoya, F. Jelezko, S. Onoda, H. Sumiya, V. Khemani, C. Von Keyserlingk, N. Y. Yao, E. Demler, and M. D. Lukin, Observation of Discrete Time-Crystalline Order in a Disordered Dipolar Many-Body System, Nature (London) 543, 221 (2017).
\bibitem{ExpDTC_StarShape_NMR_PRL} S. Pal, N. Nishad, T. S. Mahesh, and G. J. Sreejith, Temporal Order in Periodically Driven Spins in Star-Shaped Clusters, Phys. Rev. Lett. 120, 180602 (2018).
\bibitem{ObservDTC_NMR_PRL} J. Rovny, R. L. Blum, and S. E. Barrett, Observation of
Discrete-Time-Crystal Signatures in an Ordered Dipolar Many-Body System, Phys. Rev. Lett. 120, 180603 (2018).
\bibitem{ObservQuasiTC_PRL} S. Autti, V. B. Eltsov, and G. E. Volovik, Observation of a Time Quasicrystal and Its Transition to a Superfluid Time Crystal, Phys. Rev. Lett. 120, 215301 (2018).
\bibitem{ObservSTC_SuperfluidQGas_PRL}  J. Smits, L. Liao, H. T. C. Stoof, and P. van der Straten, Observation of a Space-Time Crystal in a Superfluid Quantum Gas, Phys. Rev. Lett. 121, 185301 (2018).
\bibitem{DTC_CQED_PRL} Z. Gong, R. Hamazaki, and M. Ueda, Discrete Time-Crystalline Order in Cavity and Circuit QED Systems, Phys. Rev. Lett. 120, 040404 (2018).
\bibitem{ObservDissipativeTC_PRL} H. Keßler, P. Kongkhambut, C. Georges, L. Mathey, J. G. Cosme, and A. Hemmerich, Observation of a Dissipative Time Crystal, Phys. Rev. Lett. 127, 043602 (2021).
\bibitem{DTC_DissipativeESR_NJP} J. O’Sullivan, O. Lunt, Ch. W. Zollitsch, M. L. W. T., J. J. L. Morton, and A. Pal, Signatures of Discrete Time Crystalline
Order in Dissipative Spin Ensembles, New J. Phys. 22, 085001 (2020).
\bibitem{FPre_NMR_NaturePhys} P. Peng, C. Yin, X. Huang, C. Ramanathan, and P. Cappellaro, Floquet Prethermalization in Dipolar Spin Chains, Nat. Phys. 17, 444 (2021).
\bibitem{ProgramDTC_NISQ_arxiv2007_11602} M. Ippoliti, K. Kechedzhi, R. Moessner, S. L. Sondhi, and V. Khemani, Many-Body Physics in the NISQ Era: Quantum Programming a Discrete Time Crystal, arXiv preprint arXiv:2007.11602 (2020).
\bibitem{DTC_SupercondQC_arxiv2105_06632} P. Frey and S. Rachel, Simulating a Discrete Time Crystal over 57 Qubits on a Quantum Computer, arXiv preprint arXiv:2105.06632 (2021).
\bibitem{DTC_ProgNV_arxiv2107_00736} J. Randall, C. E. Bradley, F. V. van der Gronden, A. Galicia, M. H. Abobeih, M. Markham,  D. J. Twitchen, F. Machado,  N. Y. Yao, and T. H. Taminiau, Observation of a Many-Body-Localized Discrete Time Crystal with a Programmable Spin-Based Quantum Simulator, arXiv preprint arXiv:2107.00736 (2021).
\bibitem{FPrePhase_SC_Pan_arxiv2107_07311} C. Ying, Q. Guo, S. Li, M. Gong, X.-H. Deng, F. Chen, C. Zha, Y. Ye, C. Wang, Q. Zhu, S. Wang, Y. Zhao, H. Qian, S. Guo, Y. Wu, H. Rong, H. Deng, F. Liang, J. Lin, Y. Xu, C.-Z. Peng, C.-Y. Lu, Z.-Q. Yin, X. Zhu, and J.-W. Pan, Floquet Prethermal Phase Protected by U(1) Symmetry on a Superconducting Quantum Processor, arXiv preprint arXiv:2107.07311 (2021).
\bibitem{ObservPreDTC_TrapIon_Science} A. Kyprianidis, F. Machado, W. Morong, P. Becker, K. S. Collins, D. V. Else, L. Feng, P. W. Hess, C. Nayak, G. Pagano, N. Y. Yao, and C. Monroe, Observation of a Prethermal Discrete Time Crystal, Science 372, 1192 (2021).
\bibitem{Waldher2014} G. Waldherr, Y. Wang, S. Zaiser, M. Jamali, T. Schulte-Herbrüggen, H. Abe, T. Ohshima, J. Isoya, J. F. Du, P. Neumann, and J. Wrachtrup, Quantum Error Correction in a Solid-State Hybrid Spin Register, Nature 506, 204 (2014).
\bibitem{Zaiser2016} S. Zaiser, T. Rendler, I. Jakobi, T. Wolf, S.-Y. Lee, S. Wagner, V. Bergholm, T. Schulte-Herbrüggen, P. Neumann, and J. Wrachtrup, Enhancing Quantum Sensing Sensitivity by a Quantum Memory, Nat. Commun. 7, 12279 (2016).
\bibitem{QFT_Vadim} V. Vorobyov, S. Zaiser, N. Abt, J. Meinel, D. Dasari, P. Neumann, and J. Wrachtrup, Quantum Fourier Transform for Quantum Sensing, arXiv preprint arXiv:2008.09716 (2020).
\bibitem{tommy}T. Oeckinghaus etal., Spin–Phonon Interfaces in Coupled Nanomechanical Cantilevers, Nano Lett., 20, 1, 463-469 (2020).
\bibitem{durgaprl}D. D. Bhaktavatsala Rao, Nir Bar-Gill, and Gershon Kurizki, Generation of Macroscopic Superpositions of Quantum States by Linear Coupling to a Bath, Phys. Rev. Lett. 106, 010404 (2011).
\bibitem{nvnv}F. Dolde, I. Jakobi, B. Naydenov, N. Zhao, S. Pezzagna, C. Trautmann, J. Meijer, P. Neumann, F. Jelezko, and J. Wrachtrup, Room-Temperature Entanglement between Single Defect Spins in Diamond, Nature Physics volume 9, pages139–143 (2013).
\bibitem{EntangSwapOptimalCont_NV_NC} F. Dolde, V. Bergholm, Y. Wang, I. Jakobi, B. Naydenov, S. Pezzagna, J. Meijer, F. Jelezko, P. Neumann, T. Schulte-Herbrüggen, J. Biamonte, and J. Wrachtrup, High-Fidelity Spin Entanglement Using Optimal Control, Nat. Commun. 5, 3371 (2014).

\end{thebibliography}
\end{document}